\documentclass[twocolumn]{aastex7}

\usepackage{amsmath}
\usepackage{siunitx}
\usepackage{booktabs} 
\usepackage{graphicx}
\usepackage[symbol]{footmisc}

\newcommand{\kms}{\,\ensuremath{\rm{km\,s}^{-1}}}
\newcommand{\msun}{\mbox{${\rm M}_\odot$}}
\newcommand{\mstar}{\mbox{${M}_{\rm star}$}}
\newcommand{\mhi}{\mbox{${M}_{\mathrm{H}\;\textsc{i}}$}}
\newcommand{\lya}{\mbox{${\rm Ly}\alpha$}}

\received{2025 August 12}
\revised{2025 September 16}
\accepted{2025 September 28}
\submitjournal{ApJL}
\shorttitle{CUBS--MALS \ion{H}{1} Stacking} 
\shortauthors{DePalma et al.}

\begin{document}

\title{\ion{H}{1} Properties of Field Galaxies at $\bm{z\approx 0.2}$--0.6: Insights into Declining Cosmic Star Formation}

\correspondingauthor{David DePalma}
\email{ddepalma@mit.edu}

\author[0009-0003-8927-2140,gname=David,sname=DePalma]{David DePalma}
\affiliation{MIT--Kavli Institute for Astrophysics and Space Research, 77 Massachusetts Avenue, Cambridge, MA 02139, USA}
\email{ddepalma@mit.edu}

\author[0000-0001-7547-4241,gname=Neeraj,sname=Gupta]{Neeraj Gupta}
\affiliation{Inter-University Centre for Astronomy and Astrophysics, Post Bag 4, Ganeshkhind, Pune 411 007, India}
\email{ngupta@iucaa.in}

\author[0000-0001-8813-4182,gname=Hsiao-Wen,sname=Chen]{Hsiao-Wen Chen}
\affiliation{Department of Astronomy and Astrophysics, The University of Chicago, 5640 S. Ellis Avenue, Chicago, IL 60637, USA}
\email{hwchen@uchicago.edu}

\author[0000-0003-3769-9559,gname='Robert A.',sname=Simcoe]{Robert A. Simcoe}
\affiliation{MIT--Kavli Institute for Astrophysics and Space Research, 77 Massachusetts Avenue, Cambridge, MA 02139, USA}
\email{simcoe@space.mit.edu}

\author[0000-0002-3814-9666,gname=Sergei,sname=Balashev]{Sergei Balashev}
\affiliation{Independent Researcher}
\email{s.balashev@gmail.com}

\author[0000-0003-3244-0409,gname=Erin,sname=Boettcher]{Erin Boettcher}
\affiliation{Department of Astronomy, University of Maryland, College Park, MD 20742, USA}
\affiliation{X-ray Astrophysics Laboratory, NASA/GSFC, Greenbelt, MD 20771, USA}
\affiliation{Center for Research and Exploration in Space Science and Technology, NASA/GSFC, Greenbelt, MD 20771, USA}
\email{eboettch@umd.edu}

\author[0000-0001-5804-1428,gname=Sebastiano,sname=Cantalupo]{Sebastiano Cantalupo}
\affiliation{Department of Physics, University of Milan Bicocca, Piazza della Scienza 3, I-20126 Milano, Italy}
\email{sebastiano.cantalupo@unimib.it}

\author[0000-0002-8739-3163,gname='Mandy C.',sname=Chen]{Mandy C. Chen}
\affiliation{Cahill Center for Astronomy and Astrophysics, California Institute of Technology, Pasadena, CA 91125, USA}
\affiliation{The Observatories of the Carnegie Institution for Science, 813 Santa Barbara Street, Pasadena, CA 91101, USA}
\email{mandyc@caltech.edu}

\author[0000-0003-2658-7893,gname=Françoise,sname=Combes]{Françoise Combes}
\affiliation{Observatoire de Paris, Coll\`ege de France, PSL University, Sorbonne University, CNRS, LUX, Paris, France}
\email{francoise.combes@obspm.fr}

\author[0000-0002-4900-6628,gname=Claude-André,sname=Faucher-Giguère]{Claude-André Faucher-Giguère}
\affiliation{CIERA and Department of Physics and Astronomy, Northwestern University, 1800 Sherman Avenue, Evanston, IL 60201, USA}
\email{cgiguere@northwestern.edu}

\author[0000-0001-9487-8583,gname='Sean D.',sname=Johnson]{Sean D. Johnson}
\affiliation{Department of Astronomy, University of Michigan, 1085 S. University, Ann Arbor, MI 48109, USA}
\email{seanjoh@umich.edu}

\author[0000-0002-0648-2704,gname=Hans-Rainer,sname=Klöckner]{Hans-Rainer Klöckner}
\affiliation{Max-Planck-Institut für Radioastronomie, Auf dem Hügel 69, 53121 Bonn, Germany}
\email{hkloeckner@mpifr-bonn.mpg.de}

\author[0000-0002-4912-9388,gname=Jens-Kristian,sname=Krogager]{Jens-Kristian Krogager}
\affiliation{Université Lyon, ENS de Lyon, CNRS, Centre de Recherche Astrophysique de Lyon UMR5574, 69230 Saint-Genis-Laval, France}
\affiliation{French-Chilean Laboratory for Astronomy, IRL 3386, CNRS and Universidad de Chile, Santiago, Chile}
\email{jens-kristian.krogager@univ-lyon1.fr}

\author[0000-0002-0311-2812,gname='Jennifer I-Hsiu',sname=Li]{Jennifer I-Hsiu Li}
\affiliation{Center for AstroPhysical Surveys, National Center for Supercomputing Applications, University of Illinois Urbana-Champaign, Urbana, IL 61801, USA}
\email{jli184@illinois.edu}

\author[0000-0003-0389-0902,gname=Sebastián,sname=López]{Sebastián López}
\affiliation{Departamento de Astronomía, Universidad de Chile, Casilla 36-D, Santiago 7550000, Chile}
\email{slopez@das.uchile.cl}

\author[0000-0002-5777-1629,gname=Pasquier,sname=Noterdaeme]{Pasquier Noterdaeme}
\affiliation{Institut d'astrophysique de Paris, CNRS-SU, UMR 7095, 98bis bd Arago, 75014 Paris, France}
\email{noterdaeme@iap.fr}

\author[gname=Patrick,sname=Petitjean]{Patrick Petitjean}
\affiliation{Institut d'astrophysique de Paris, CNRS-SU, UMR 7095, 98bis bd Arago, 75014 Paris, France}
\email{ppetitje@iap.fr}

\author[0000-0002-2941-646X,gname=Zhijie,sname=Qu]{Zhijie Qu}
\affiliation{Department of Astronomy E304, Physics Building, Tsinghua University, Beijing, China}
\email{quzhijie@tsinghua.edu.cn}

\author[0000-0002-8459-5413,gname='Gwen C.',sname=Rudie]{Gwen C. Rudie}
\affiliation{The Observatories of the Carnegie Institution for Science, 813 Santa Barbara Street, Pasadena, CA 91101, USA}
\email{gwen@carnegiescience.edu}

\author[0000-0002-0668-5560,gname=Joop,sname=Schaye]{Joop Schaye}
\affiliation{Leiden Observatory, Leiden University, PO Box 9513, 2300 RA Leiden, the Netherlands}
\email{schaye@strw.leidenuniv.nl}

\author[0000-0001-7869-2551,gname=Fakhri,sname=Zahedy]{Fakhri Zahedy}
\affiliation{Department of Physics, University of North Texas, Denton, TX 76201, USA}
\email{fakhri.zahedy@unt.edu}

\begin{abstract}

We report statistically significant detection of \ion{H}{1} 21~cm emission from intermediate-redshift ($z\approx 0.2$--0.6) galaxies.  By leveraging multi-sightline galaxy survey data from the Cosmic Ultraviolet Baryon Survey (CUBS) and deep radio observations from the MeerKAT Absorption Line Survey (MALS), we have established a sample of $\approx 6000$ spectroscopically identified galaxies in 11 distinct fields to constrain the neutral gas content at intermediate redshifts.  The galaxies sample a broad range in stellar mass, from $\log\,\mstar/\msun\approx 8$ to $\log\,\mstar/\msun\approx 11$, with a median of $\langle\log\,\mstar/\msun\rangle_{\rm med}\approx 10$ and a wide range in redshift from $z\approx 0.24$ to $z\approx 0.63$ with a median of $\langle\,z\rangle_{\rm med}=0.44$.  While no individual galaxies show detectable \ion{H}{1} emission, the emission line signal is detected in the stacked spectra of all subsamples at greater than 4-$\sigma$ significance. The observed total \ion{H}{1} 21~cm line flux translates to a \ion{H}{1} mass, \mhi\ $\approx10^{10}$\msun. We find a high \ion{H}{1}-to-stellar mass ratio of $\mhi/\mstar\approx 6$ for low-mass galaxies with $\langle\log\,\mstar/\msun\rangle\approx 9.3$ ($>3.7\,\sigma$). For galaxies with $\langle\log\,\mstar/\msun\rangle\approx 10.6$, we find $\mhi/\mstar\approx 0.3$ ($>4.7\,\sigma$).  In addition, the redshift evolution of \ion{H}{1} mass, $\langle\mhi\rangle$, in both low- and high-mass field galaxies, inferred from the stacked emission-line signal, aligns well with the expectation from the cosmic star formation history.  This suggests that the overall decline in the cosmic star formation activity across the general galaxy population may be connected to a decreasing supply of neutral hydrogen. Finally, our analysis has revealed significant 21~cm signals at distances greater than 75 kpc from these intermediate-redshift galaxies, indicating a substantial reservoir of \ion{H}{1} gas in their extended surroundings.
\end{abstract}

\section{Introduction} 
\label{sec:intro}  

Since {\it Cosmic Noon} at redshift $z\approx 2$, the cosmic star formation rate density (SFRD) has decreased by nearly a factor of ten \citep[e.g.,][]{Madau2014} and possibly larger \citep[e.g.,][]{Matthews2024}. The physical cause of such a dramatic decline remains unclear.  Neutral gas is a crucial component in maintaining star formation in galaxies \citep[e.g.,][]{Walter2020}. The observed decline in cosmic SFRD after $z=2$ may, therefore, imply a simultaneous change in the properties of neutral gas content of galaxies.

The \ion{H}{1} 21~cm line provides a direct method for measuring the neutral gas content of galaxies. However, detecting \ion{H}{1} in 21~cm line emission during the ``cosmic afternoon" beyond the nearby universe remains exceedingly challenging with current observational facilities.  Direct detections of \ion{H}{1} 21-cm emission line from individual galaxies, even at modest redshifts of $z\sim$0.2--0.3, are either based on selecting extremely massive star-forming galaxies \citep[e.g.,][]{Catinella2015} or required several hundred to a few thousand hours of total integration time in small regions of the sky \citep[e.g.,][]{Gogate2023}. As a result, the neutral hydrogen content of individual galaxies beyond $z\approx 0.2$ remains largely unexplored \citep[cf.,][]{Jarvis2025}.

\begin{deluxetable}{ccccccr}
    \tablecaption{Summary of CUBS-MALS fields\label{Table:fields}}
    \tablehead{CUBS& & \multicolumn{2}{c}{MALS pointing center}& &\multicolumn{2}{c}{Median and 68\% dispersion}\\ \cline{3-4} \cline{6-7}
    \colhead{QSO ID}& $z_{\rm QSO}^\dag$ & \colhead{R.~A.}&\colhead{Decl.}&\colhead{$N_{\rm gal}^\ddag$}&\colhead{$z_{\rm gal}$}& \colhead{$\log{\left(M_\mathrm{star}/\mathrm{M_\sun}\right)}$}}
    \startdata
    J0110$-$1648 & 0.782 & 01:10:35.13&$-$16:48:31.1&743&0.473 [0.314, 0.547] &10.2 [9.3, 10.8]\\
    J0111$-$0316 & 1.238 & 01:12:39.18&$-$03:28:43.3&361&0.424 [0.315, 0.577] &9.8 [9.0, 10.8]\\
    J0119$-$2010 & 0.816 & 01:19:45.20&$-$19:58:27.9&532&0.468 [0.350, 0.604] &10.2 [9.2, 10.8]\\
    J0154$-$0712 & 1.293 & 01:54:08.54&$-$06:52:34.1&551&0.469 [0.369, 0.590] &10.1 [9.2, 10.8]\\
    J0333$-$4192 & 1.115 & 03:31:57.66&$-$40:58:40.7&478&0.394 [0.320, 0.563] &10.0 [9.1, 10.7]\\
    J0357$-$4812 & 1.013 & 03:57:21.82&$-$48:12:14.3&419&0.432 [0.332, 0.584] &10.2 [9.2, 10.8]\\
    J0420$-$5650 & 0.948 & 04:20:47.23&$-$57:12:52.8&609&0.424 [0.269, 0.538] &9.9 [9.0, 10.7]\\
    J2135$-$5316 & 0.812 & 21:34:18.19&$-$53:35:14.7&665&0.450 [0.330, 0.605] &9.9 [9.1, 10.7]\\
    J2245$-$4931 & 1.001 & 22:45:01.25&$-$49:31:32.3&578&0.530 [0.354, 0.587] &10.1 [9.2, 10.8]\\
    J2308$-$5258 & 1.073 & 23:07:33.06&$-$53:12:58.9&487&0.413 [0.325, 0.573] &10.1 [9.2, 10.7]\\
    J2339$-$5523 & 1.354 & 23:39:13.22&$-$55:23:50.4&563&0.419 [0.294, 0.542] &9.9 [9.0, 10.7]
    \enddata
    \tablenotetext {$\dag$}{Redshift of the CUBS QSO.} {$\ddag$}{Number of CUBS galaxies in the field.}
\end{deluxetable}

In contrast, stacking techniques that combine radio maps of 
galaxies with known redshifts from optical surveys have proven effective in extracting statistically significant \ion{H}{1} signals in galaxy populations up to $z\approx 1$ \citep[e.g.,][]{Chowdhury2020}. At somewhat lower redshifts ($z\approx 0.2$--0.5), stacking measurements based on star-forming galaxies in the COSMOS field \citep[][]{Scoville2007} and the Extended Groth Strip \citep[EGS;][]{Newman2013} have successfully revealed the mean \ion{H}{1} 21-cm emission signal from these galaxies \citep[see results from the uGMRT, MIGHTEE and CHILES surveys; e.g.,][]{Bera2023, Sinigaglia2024MIGHTEE-HExperiment, Bianchetti2025NewData}. However, these measurements exhibit significant scatter, likely due to sample selection effects and variations between different fields caused by cosmic variance \citep[][their Fig.~7]{Bianchetti2025NewData}.  It is therefore necessary to construct a uniform galaxy sample covering vastly different locations in the cosmos to address possible field-to-field variations.

In this letter, we report the initial findings of an analysis of stacking \ion{H}{1} using deep galaxy survey data from the Cosmic Ultraviolet Baryon Survey \citep[CUBS;][]{Chen2020The1} and associated radio observations from the MeerKAT Absorption Line Survey \citep[MALS;][]{Gupta2016TheMALS}.  Across eleven cosmologically distinct fields, we have assembled a random sample of $\approx 6000$ spectroscopically identified galaxies at $0.24\leq z\leq0.63$ with sensitive \ion{H}{1} coverage available from the MALS program. We report new constraints on the \ion{H}{1} content of these intermediate redshift galaxies.  The large galaxy sample has also enabled a detailed study of the redshift evolution and stellar mass dependence of \ion{H}{1} mass across these 11 fields.  Throughout this work, we adopt a flat $\Lambda$ cosmology with $H_0=\qty{70}{km.s^{-1}.{Mpc}^{-1}}$ and $\Omega_{\rm M,0}=0.3$.

\section{The Galaxy Sample}
\label{sec:samp}

To obtain a robust constraint on the \ion{H}{1} content of the distant galaxy population, we explore the synergy between the CUBS galaxy sample and the MALS program, one of the early key science programs on the MeerKAT telescope \citep{Jonas2016TheTelescope}.  A summary of galaxy fields with available galaxy survey data and MALS observations is presented in Table \ref{Table:fields}. Here, we provide a brief summary of the main characteristics of the two programs.

\subsection{The CUBS galaxy sample}
\label{sec:cubs}

CUBS is a large {\it Hubble Space Telescope} ({\it HST}) Cycle 25 General Observer Program (GO-CUBS; ${\rm PID}=15163$; PI: Chen), designed to map diffuse baryonic structures at $z\lesssim 1$ using quasar absorption spectroscopy in 15 UV-bright quasar fields \citep[][]{Chen2020The1}.  A critical component of the CUBS program is a comprehensive deep-galaxy survey in these 15 fields, covering an area of angular radius $\theta\lesssim10\arcmin$ from each QSO sightline.  The galaxy spectroscopic survey aims to reach a uniformly high completeness for $L_*$-type galaxies across each QSO field with progressively deeper depth to uncover fainter galaxies, $\lesssim 0.1\,L_*$, closer to the quasar sightline.  This tiered survey design is motivated by the scientific objectives of probing individual gaseous halos on scales of \qtyrange{10}{100}{kpc} using quasar absorption spectroscopy while characterizing the large-scale galaxy clustering environments on scales of \qtyrange{1}{5}{Mpc}.  

\begin{figure*}
    \plotone{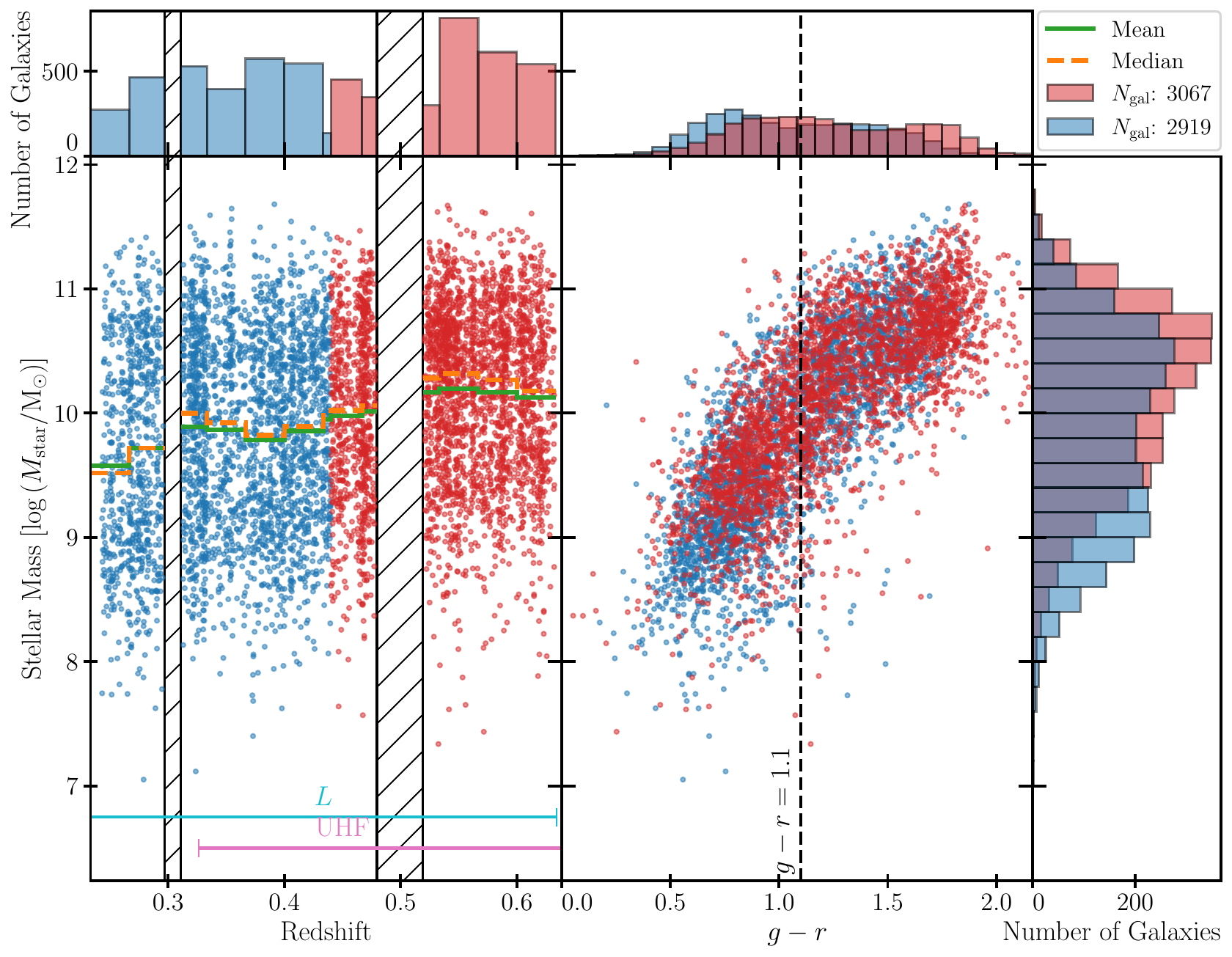}
    \caption{Summary of galaxy properties in the CUBS-MALS galaxy sample. 
    The data points are color-coded according to redshift, with galaxies at $z<0.44$ shown in blue and higher-redshift galaxies in red.  The redshift distribution of stellar mass is presented in the {\it left} panels.
    In the bottom panel, the blue and red lines indicate the corresponding redshift coverage of the $L$ and UHF bands, respectively. Hatched regions demarcate bands of RFI. Except where truncated by RFI or the samples' boundary, the bins in the histogram of redshifts in the top panel bridge multiples of $\Delta z=1/30$. The dashed orange and solid green stairs represent the moving weighted median and weighted mean, respectively (see \S\ref{sec:stacking} for details), within each bin of redshift. The stellar mass versus the observed $g-r$ color is displayed in the {\it middle} panels with the $g-r=1.1$ color threshold marked by a dashed line. In the histogram of stellar mass in the {\it right} panel, the bins span multiples of $0.2\;\mathrm{dex}$. 
    The CUBS-MALS galaxy sample spans a broad range in stellar mass, from $\log\,\mstar/\msun\approx 8$ to $\log\,\mstar/\msun\approx 11$, with a median of $\langle\log\,\mstar/\msun\rangle_{\rm med}\approx 10$ and a wide range in redshift from $z\approx 0.24$ to $z\approx 0.63$ with a median of $\langle\,z\rangle_{\rm med}=0.44$.  The observed $g-r$ color correlates well with \mstar\ but with a large scatter. \label{fig:RedshiftStellarMass}}
\end{figure*}

To optimize the survey efficiency, the wide-field survey prioritizes spectroscopic observations for massive galaxies that are expected to be bright and red with $AB(I)\approx 22.5$ mag and optical colors of $g-r\gtrsim 1.1$ \citep[e.g.,][see also Figure \ref{fig:RedshiftStellarMass} below]{Johnson2015}.  At $z=0.8$, this survey depth corresponds to $M_{B}=-20.6$ for a typical spiral galaxy, in comparison to the characteristic absolute magnitude of $M_{B_*}\approx -21$ from random galaxy surveys \citep[see e.g.,][]{Faber2007,Cool2012}. Fainter and bluer galaxies are included as fillers for the slit mask design and are therefore covered at a lower completeness level.  Within the inner $3\arcmin$ region around the QSO, the survey aims to obtain a magnitude-limited sample with $AB(r)>24$ mag, regardless of the intrinsic color. Redshift measurements are first based on cross-correlation with the Sloan Digital Sky Survey galaxy templates \citep[e.g.,][]{Chen2009, Johnson2013}, and then visually inspected by two independent observers. Uncertainties in the redshift measurements are typically $\Delta\,v\approx\qty{40}{\km\per\s}$. A complete description and characterization of the galaxy sample will be presented in a future paper (Johnson et al., in preparation). In short, for each spectroscopically identified galaxy, we estimate the underlying stellar mass, \mstar, based on the observed broad-band spectral energy distribution using the Bayesian Analysis of Galaxies for Physical Inference and Parameter EStimation code \citep[BAGPIPES;][]{Carnall2018} and a stellar initial mass function (IMF) from \cite{Kroupa2002}. Uncertainties in \mstar\ are typically 0.2 dex, primarily driven by the adopted stellar IMF (DePalma et al., in preparation).  Eleven of the 15 CUBS QSO fields are also covered by the MALS program.  The CUBS--MALS QSO fields are summarized in Table \ref{Table:fields}.

\subsection{MALS observations}
\label{sec:mals}

MALS is a large program on the MeerKAT telescope, designed to carry out the most sensitive search of \ion{H}{1} 21~\unit{\cm} and OH \num{18}-\unit{\cm} absorption lines at $z<2$ to establish reliable measurements of the evolution of cold atomic and molecular gas cross-sections of galaxies \citep[][]{Gupta2016TheMALS}. The survey covers about 400 pointings centered at radio sources brighter than \qty{200}{mJy} at \qty{1}{\GHz} in the $L$-band, from 900 to 1670 MHz, and the UHF band, from 580 to 1015 MHz. Simultaneously, MeerKAT telescope's large field of view with a full width at half maximum (FWHM) of $\theta\approx88\arcmin$ at \qty{\sim1}{\GHz} and excellent sensitivity render it a highly competitive survey for \ion{H}{1} 21-\unit{\cm} emission line \citep[e.g.,][]{Boettcher21} and radio-continuum signals \citep[e.g.,][]{Deka24dr1, Wagenveld24}. 

All MALS data are processed using the Automated Radio Telescope Imaging Pipeline \citep[ARTIP;][]{Gupta21} based on NRAO's Common Astronomy Software Applications (CASA) package \citep[][]{TheCASATeam2022CASAAstronomy}. Detailed descriptions regarding the processing and calibrations of the $L$-band and UHF band data are presented in \citet{Gupta21} and \cite{Combes21}.  For spectral-line imaging, the continuous frequency band after calibration is divided into 15 overlapping spectral windows (SPWs) labeled as SPW 0 to SPW 14. Continuum-subtracted visibility datasets for each SPW are processed to produce Stokes-$I$ spectral line cubes, which are corrected for the heliocentric motion of the Earth.  The corrections for the  primary-beam attenuation are applied afterwards.  

For spectral stacks, we focus on galaxies at $0.24<z<0.63$ across the 11 CUBS-MALS fields, where most sensitive constraints for the corresponding 21~\unit{cm} line are possible with the MALS data.  These include 121 spectral-line cubes from SPWs 0--4 in the $L$ band and SPWs 9--14 in UHF. The cubes have a spatial extent of 3K$\times$3K pixels with a pixel size of $2\arcsec$ in the $L$ or $3\arcsec$ in the UHF band. The typical spatial resolution, as determined by the full width at half maximum (FWHM) of the synthesized beam, is $11\arcsec$ at 1100 MHz (corresponding to the observed frequency of the 21~cm line at $z=0.3$) and $22\arcsec$ at 900 MHz (corresponding to the observed frequency of the 21~cm line at $z=0.6$).  The frequency channel spacings for $L$- and UHF-band cubes are \qty{26.123}{\kHz} and \qty{16.602}{\kHz}, respectively, corresponding to \qty{7.8}{\kms} and \qty{5.0}{\kms} at \qty{1}{\GHz}.

The number of galaxies included in the analysis from each field, as well as median redshift and median \mstar, are presented in Table \ref{Table:fields}. Figure \ref{fig:RedshiftStellarMass} presents the redshift distribution of \mstar\ ({\it left} panel) and \mstar\ versus the observed $g-r$ color ({\it middle} panel) of the current CUBS--MALS galaxy sample.  
Note that due to significant radio frequency interference (RFI) in certain frequency ranges the spectral coverage is effectively not continuous.  
In summary, the full sample comprises 5986 galaxies with $\log\,\mstar/\msun\approx 7.1$--$11.7$ and a median redshift of $\langle z\rangle_{\rm med}\approx 0.44$.  The observed $g-r$ color spans a broad range, from $g-r<0.5$ to $g-r\approx 2$, and correlates strongly with \mstar.  However, the scatter is large, suggesting that additional physical properties may be driving the observed $g-r$ color \citep[e.g.,][]{Bell2003}.
To explore how the mean \ion{H}{1} content varies with redshift and galaxy properties, we further divide the galaxy sample into four subsamples of low-mass ($\log\,\mstar/\msun<10$) and high-mass ($\log\,\mstar/\msun\ge 10$) at $z<0.44$, and low mass and high mass at $z>0.44$.  See Table \ref{Table:Samples} for details.

\section{Stacking Analysis}
\label{sec:stacking}

To measure the mean strength of the 21-cm line signal in intermediate-redshift galaxies, we stack MALS spectrum at the observed frequency of the redshifted 21-cm line of each galaxy following the steps described here.  First, we correct each cube for its primary beam's attenuation using the {\tt katbeam} (version 0.1) model \citep{Mauch20}. Then, for each galaxy, we extract spectral cylinders centered on its systemic redshift with a line-of-sight velocity width of $\pm\qty{1000}{\km\per\s}$, and position with a range in diameter from 150 to 600 kpc.  Because the size of the synthesized beam is frequency-dependent and the corresponding physical scale per beam also varies with galaxy redshift, we experiment with multiple aperture sizes when extracting the spectra. Specifically, at redshift $z = 0.634$, a FWHM of \qty{22}{\arcsec} corresponds to roughly 150 kpc in proper distance, representing the coarsest spatial resolution in our MALS dataset. Therefore, we begin with an aperture diameter of 150 kpc---sufficiently large to encompass the expected \ion{H}{1} envelope of these galaxies \citep[e.g.,][]{Bosma2017,Wang2024}---and increase the aperture size in steps of 150 kpc up to a maximum diameter of 600 kpc.
In Section \ref{sec:diameter}, we demonstrate that adopting a physical aperture size of 300 kpc delivers the highest quality signals in the stacked spectra.

Next, along the spectral dimension, we subtract residual continuum from each cylinder by fitting a first-order polynomial, excluding the central $\pm\qty{250}{\km\per\s}$ window. 
Note that the radio continuum emission has already been subtracted from the visibilities before making spectral line cubes.
Then, we integrate the cylinder over its spatial dimensions to obtain a one-dimensional spectrum in flux density units.
%
To combine individual one-dimensional spectra, we resample each spectrum onto a common velocity grid with a bin size of $\Delta v = \qty{50}{\km\per\s}$ and zero velocity corresponding to the systemic redshift of the galaxy.  We then coadd individual spectra, $S_i(v)$ of galaxy $i$, over an ensemble of $N$ galaxies and compute both a weighted mean $\langle S(v) \rangle_{\rm avg}$ and median $\langle S(v) \rangle_{\rm med}$.  
Because the root-mean-square (RMS) noise in the extracted MALS spectra depends on the primary beam attenuation and the frequency-dependent gain variation of the telescope, the noise between galaxies' spectra varies by a factor of $\approx 2$--3.  
We therefore apply inverse-variance weighting for each galaxy, where the weight is $w_i= \sigma_i^{-2}/\sum_{i=1}^N \sigma_i^{-2}$ with $\sigma_i$ being the RMS noise in the extracted MALS spectrum.  For galaxies with both $L$ and UHF coverage, we obtain an average of the $L$ and UHF spectra, applying the same inverse-variance weighting before incorporating these galaxies into the full stack.

We convert the stacked flux spectrum to \ion{H}{1} mass spectrum following,
\begin{multline}
    \label{eq:himass}
    \Biggl \langle \frac{M_{\mathrm{H}\,\textsc{i}}(v)}{\mathrm{M_\sun}} \Biggr \rangle =\\ \left(\frac{2.35\times10^{5}\mathrm{}}{1+z}\right)\left(\frac{D_L(z)}{\mathrm{Mpc}}\right)^2\left(\frac{\langle S(v) \rangle}{\mathrm{Jy\,\kms}}\right),
\end{multline}
where $D_L$ is the luminosity distance computed at either the weighted mean or median redshift of each galaxy ensemble \citep[e.g.,][]{Meyer2017}.  The coefficient $2.35\times 10^5$ converts 21~cm photon velocity-integrated fluxes in Jy\,\kms\ to \ion{H}{1} mass in units of \msun.
%
In total, we include 5986 unique galaxies at $0.24<z<0.63$, 4783 of which are observed in both $L$ and UHF, leading to 10,887 individual spectra from the $L$- or UHF-band for stacking.

\begin{figure*}
    \plottwo{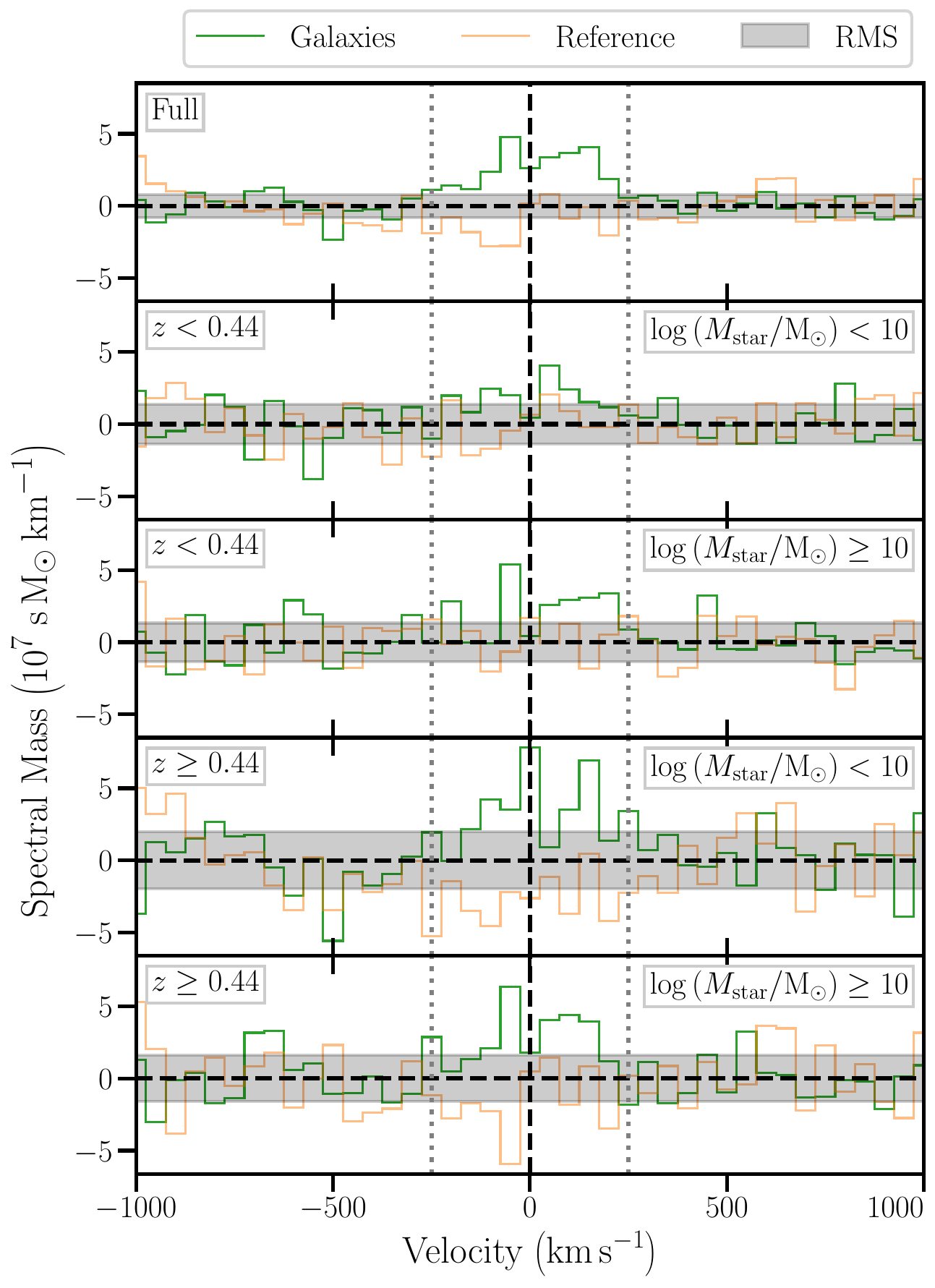}{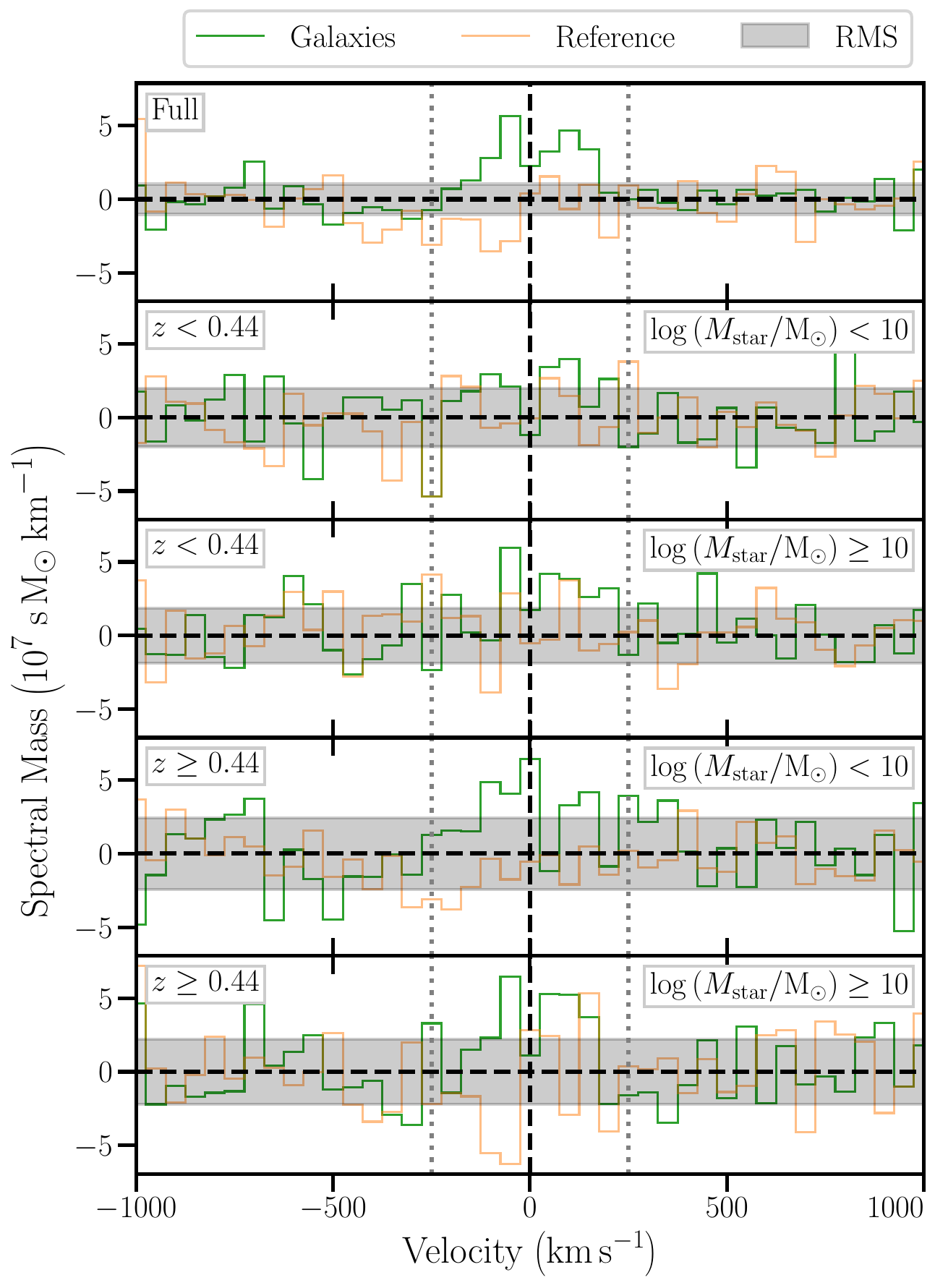}
    \caption{Weighted mean ({\it left}) and weighted median ({\it right}) 
    spectra of the \ion{H}{1} 21~cm emission line extracted using a physical aperture of 300 kpc in diameter around CUBS-MALS galaxies at $0.24<z<0.63$.  The stacked spectra of CUBS galaxies are shown in thick green, while the reference spectra are shown in thin orange (see \S\ \ref{sec:stacking} for details).  The gray band marks the RMS fluctuations in the error spectrum.  The signal associated with the full sample is presented in the top panel, where the signals extracted for individual subsamples are presented in the bottom four rows.
    The dotted vertical lines mark the velocity window for extracting the total \ion{H}{1} mass for each sample. The 21~cm line is confidently detected in all samples with high-$z$ galaxies displaying the strongest signal strength in both low- and high-mass galaxies (bottom two panels).
    The mean stacks for 150\,kpc- and 450\,kpc-diameter apertures are shown in Appendix~\ref{sec:Stacks-app}.
    }
    \label{Fig:Stacks}
\end{figure*}

%
For each galaxy, we also generate a corresponding reference cylinder at the same redshift but positioned at a randomly chosen location with a sufficiently large offset, ranging from the adopted aperture size to $\approx 4'$, from the science target to prevent blending with the target galaxy. If the randomly selected position overlaps with any known galaxy in the sample, we redraw until a suitable, isolated position is found. These reference cylinders undergo the same processing steps as the science targets, including the extraction of one-dimensional spectra. The resulting spectra are stacked to produce a composite reference spectrum, which serves as a baseline for comparison with the science targets. The results for both the full sample and various subsamples, along with the corresponding reference spectra, are presented in Figure \ref{Fig:Stacks}.

As shown in Figure \ref{Fig:Stacks}, the weighted mean and weighted median stacks yield consistent signal strengths, although the weighted median exhibits larger noise fluctuations. Therefore, we focus on the weighted mean spectra in the subsequent analysis.  In all subsamples, the stacked 21~cm signals span a broad velocity window from $-250$ \kms\ to $+250$ \kms, similar to what has been seen in previous measurements \citep[e.g.,][]{Chowdhury2020, Guo2021, Bianchetti2025NewData}. As described in \S\ \ref{sec:cubs}, the redshift measurements of CUBS galaxies have uncertainties of only $\Delta\,v\approx 40$ \kms, well within the spectral bin size.   Therefore, the large velocity width of the 21~cm line cannot be attributed to redshift uncertainties.  At the same time, galaxies of $\log\,\mstar/\msun=11$ typically reside in dark matter halos of mass $\log\,M_h/\msun\approx 13$ \citep[e.g.,][]{Behroozi2019} with a corresponding projected line-of-sight virial velocity of $\approx 200$ \kms\ at $z=0.6$.  The observed line width suggests a significant contribution from more spatially extended gas in the outskirts of gaseous disks and beyond \citep[see e.g.,][]{Wang2024}.  

After obtaining the stacked spectrum in mass, we integrate over a velocity window to determine the total \ion{H}{1} mass.  By experimenting with different velocity widths, we find that a window of $\pm\qty{250}{\km\per\s}$ captures the full 21~cm line flux while maximizing the significance of the total integrated \ion{H}{1} mass.

\section{Discussion and Conclusions}
\label{sec:res}

By leveraging multi-sightline galaxy survey data from CUBS and deep radio observations from MALS, we have established a statistically significant galaxy sample from 11 distinct fields to constrain the properties of \ion{H}{1} gas at intermediate redshifts (see Figure \ref{fig:RedshiftStellarMass} for a summary of the CUBS-MALS galaxy sample).  While no individual galaxies show detectable \ion{H}{1} signals, Figure \ref{Fig:Stacks} demonstrates that the \ion{H}{1} emission line is detected in the stacked spectra of all subsamples at greater than 4-$\sigma$ significance. The observed total 21~cm line flux translates to significant \ion{H}{1} mass, which is summarized in Table \ref{Table:Samples} for individual samples.  Uncertainties in $\langle\,M_{\rm HI}\rangle$ are estimated based on the RMS noise in the mass spectrum over 500\,\kms\ outside of the measurement window. Here, we discuss the significance and implications for the properties of \ion{H}{1} gas in field galaxies at intermediate redshifts.

\subsection{Optimal Extraction Apertures and the spatial extent of \ion{H}{1}}
\label{sec:diameter}

\begin{figure}
    \includegraphics[width=0.45\textwidth]{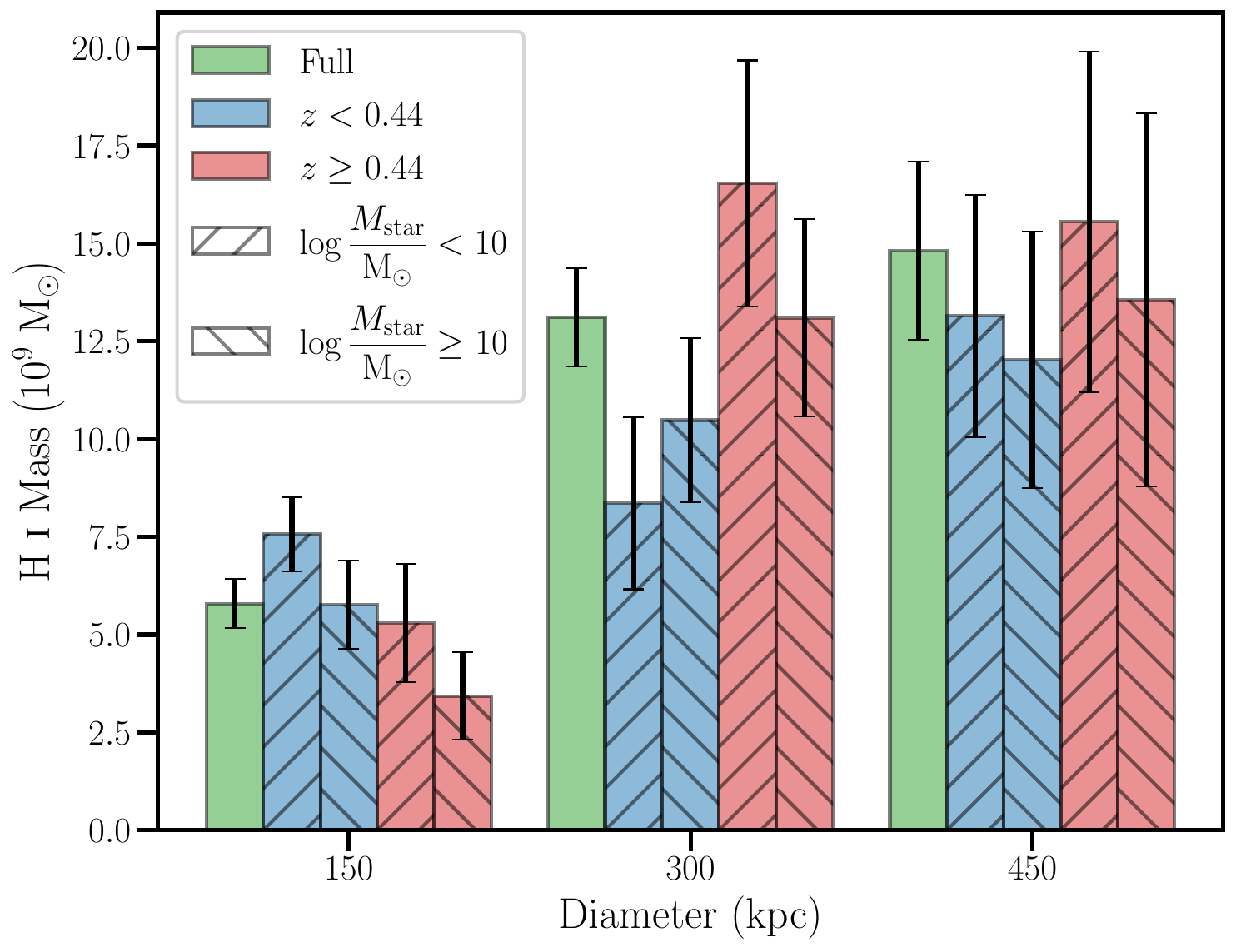}
    \caption{Cumulative mean \ion{H}{1} mass with increasing aperture diameter. Doubling the extraction aperture diameter, from 150 kpc to 300 kpc, increases the extracted total \ion{H}{1} 21-cm line flux by more than a factor of two, indicating a significant amount of neutral gas located beyond 75 kpc from the central galaxies, while further increasing the aperture diameter to 450 kpc results in a substantial increase in noise without a corresponding gain in signal (see also Figure \ref{Fig:Stacks-app}).}
    \label{fig:Diameter}
\end{figure}

The \ion{H}{1} extent of galaxies beyond the local Universe is unknown. Previous studies have adopted an extraction aperture either comparable to the size of the telescope synthesized beam \citep[e.g.][]{Bera2023} or larger \citep[$3\times$; e.g.,][]{Sinigaglia2024MIGHTEE-HExperiment,Bianchetti2025NewData}.  This approach is not suitable for the CUBS-MALS sample due to its broad redshift coverage, spanning from $z \approx 0.24$ to $z \approx 0.63$. A fixed angular size would translate into nearly a factor-of-two variation in physical scale between the lowest- and highest-redshift galaxies in our sample, resulting in inconsistent physical volumes probed for \ion{H}{1} signals across different epochs \citep[see also][]{Jones2016}.
To address this caveat, we define the extraction aperture size in physical units at the rest frame of each galaxy and select an aperture large enough to enclose the majority of the signal yet sufficiently constrained to preserve the significance of the \ion{H}{1} detections around our galaxies.

We examine the stacked \ion{H}{1} signal over a range of aperture sizes in multiples of $\sim$\qty{150}{kpc} in diameter, the largest proper distance corresponding to the beam FWHM---\qty{22}{\arcsec} at maximal redshift $z=0.634$ at \qty{869}{\MHz} among all the fields and the SPWs. We note that an aperture of diameter 150 kpc (75 kpc in radius) is already sufficiently large to cover the largest \ion{H}{1} halos known in the local universe \citep[e.g.,][]{Wang2024}.

This varying aperture exercise has yielded notable findings. While the 150 kpc-diameter aperture produces the highest SNR in the stacked spectra across all subsamples, the 300 kpc-diameter stacked spectrum shows a \ion{H}{1} 21~cm line flux more than twice as strong as that measured within the smaller aperture. However, further increasing the aperture size to 450 kpc and beyond provides only marginal gains in the measured \ion{H}{1} line flux, as noise begins to dominate (Figure \ref{fig:Diameter}; see also Figure \ref{Fig:Stacks-app}). We therefore conclude that an aperture diameter of 300 kpc (radius of 150 kpc) is optimal for fully sampling the synthesized beam and for robustly measuring the total \ion{H}{1} signal from galaxies across the entire CUBS-MALS redshift range.

The observed increase in the integrated 21-cm line flux when enlarging the aperture from 150 kpc- to 300 kpc-diameter indicates the presence of significant amounts of \ion{H}{1} gas at distances greater than 75 kpc from the CUBS-MALS galaxies.  However, because doubling the aperture diameter quadruples the enclosed area, this flux increase corresponds to a factor of $\approx 3$ decline in the mean \ion{H}{1} surface mass density, from the inner halo (within 75 kpc) to the outskirts (between 75 and 150 kpc).  Such extended reservoirs of \ion{H}{1} gas may arise from satellite galaxies and their interactions with central galaxies \citep[e.g.,][]{deBlok2018,Chen2019,Wang2023}, or may reflect rapidly cooling and turbulent gaseous halos \citep[e.g.,][]{Stern2021}.

\begin{table*}
\centering
\caption{Summary of \ion{H}{1} properties for different CUBS-MALS subsamples$^\P$} 
\addtolength{\tabcolsep}{-2pt}
\footnotesize
\begin{tabular}{lcrccccc}
\toprule
    \multicolumn{1}{c}{Sample} & $z_{\rm gal}$ & $\log\left(M_\mathrm{star}/\mathrm{M_\sun}\right)$ & $N_{\rm gal}$&$\langle z\rangle$&$\langle\log\,M_\mathrm{star}/\mathrm{M_\sun}\rangle$&
    $\langle M_{\mathrm{H}\;\textsc{i}}\rangle/{10}^9\;\mathrm{M_\sun}$ & $M_{\rm HI}/\mstar$ \\
\midrule
    Full & [0.24, 0.63]& [7.1, 11.7]& 5986 & $0.50\pm0.09$ & $10.0\pm0.7$ & $13.1\pm1.3$ & $1.3\pm 0.1$\\
    low-$z$, low-\mstar\ & [0.24, 0.44]& [7.1, 10.0] & 1319 & $0.38\pm0.04$ & $9.2\pm0.5$ & $8.4\pm2.2$ & $5.3\pm 1.4$ \\
    low-$z$, high-\mstar\ & [0.24, 0.44] & [10.0, 11.7] & 1600 & $0.38\pm0.04$ & $10.5\pm0.4$ & $10.5\pm2.1$ & $0.33\pm 0.07$\\
    high-$z$, low-\mstar\ & [0.44, 0.63]& [7.1, 10.0] & 1827 & $0.55\pm0.06$ & $9.4\pm0.4$ & $16.5\pm3.1$ & $6.6\pm 1.2$\\
    high-$z$, high-\mstar\ & [0.44, 0.63] & [10.0, 11.7] & 1240 & $0.55\pm0.05$ & $10.6\pm0.4$ & $13.1\pm2.5$ & $0.33\pm 0.06$\\
\bottomrule
\multicolumn{8}{@{}p{0.85\textwidth}}{$^\P$Errors in $\langle z \rangle$ and $\langle \log\,\mstar/\msun \rangle$ represent the standard deviation of these quantities within each sample, while errors in $\langle\,M_{\rm HI}\rangle$ represent the 1-$\sigma$ uncertainties based on the RMS noise beyond the measurement window of $\pm\qty{250}{\km\per\s}$.}
\end{tabular}
    \label{Table:Samples}
\end{table*}

\subsection{Evolution of \mhi\ with redshift}
\label{sec:zevol}

\begin{figure*}
\centering
    \includegraphics[width=0.475\textwidth]{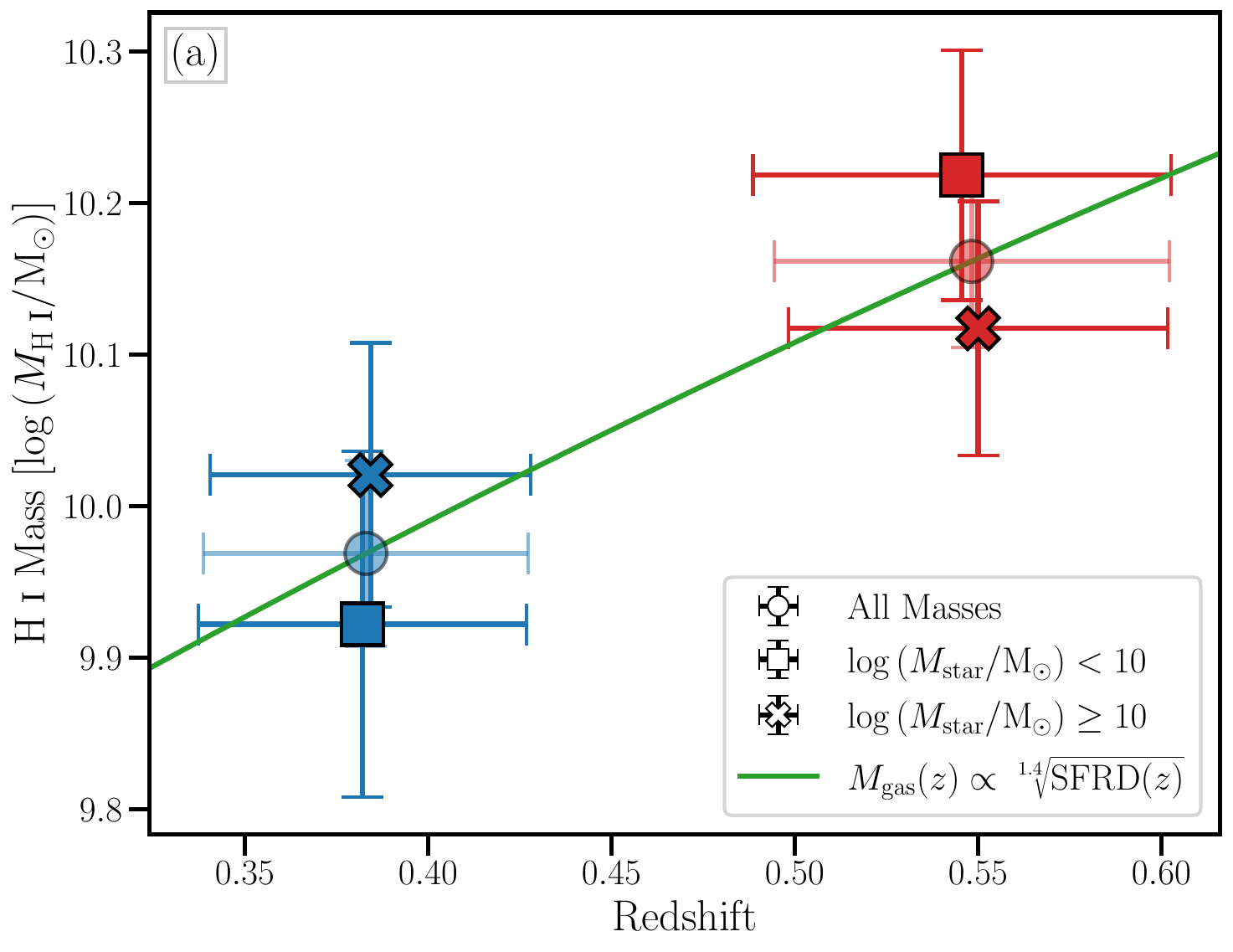}
    \includegraphics[width=0.475\textwidth]{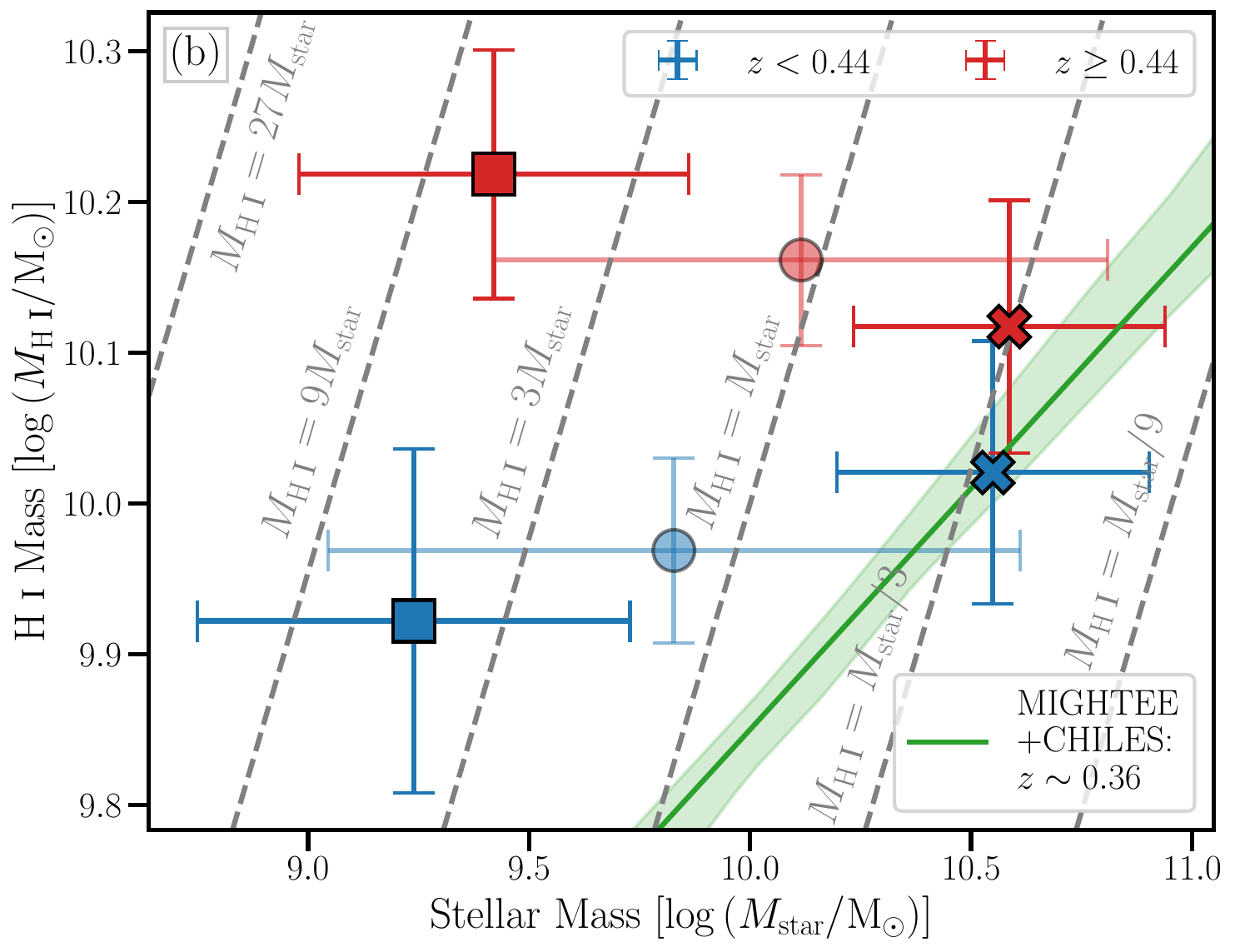}
    \caption{Variation of mean \ion{H}{1} mass with redshift (panel a) and galaxy stellar mass (panel b).  Data points and vertical error bars mark the weighted mean \ion{H}{1} mass measurements and the associated uncertainties, color-coded according to redshift following what is presented in Figure \ref{fig:RedshiftStellarMass}.  The horizontal bars indicate the dispersion of each subsample. For comparison, the solid green curve in panel (a) shows the anticipated redshift evolution of neutral gas content inferred from the cosmic star-formation rate density \citep[][Equation 15]{Madau2014}, assuming a Kennicutt--Schmidt relation \citep[][see \S\ \ref{sec:zevol} for details]{Kennicutt1998}.  The solid green line and the associated shaded band in panel (b) show the scaling relation and uncertainties from the combined MIGHTEE-\ion{H}{1}+CHILES study by \cite{Bianchetti2025NewData}. The gray dashed lines mark fixed \mhi/\mstar\ ratios to facilitate visual calibrations.}
\label{Fig:Subsamples}
\end{figure*}

Following the measurements in Table~\ref{Table:Samples}, Figure \ref{Fig:Subsamples}a shows the weighted mean \mhi\ versus redshift for different mass bins.  For comparison, we also include the anticipated redshift evolution inferred from the observed cosmic SFRD from \citet{Madau2014}, using the Kennicutt--Schmidt relation to connect star formation with neutral gas \citep[e.g.,][]{Kennicutt1998}.  

Specifically, the Kennicutt--Schmidt relation states that the star formation rate per unit area ($\Sigma_{\rm SFR}$) is proportional to the neutral gas surface mass density ($\Sigma_{\rm gas}$) according to $\Sigma_{\rm SFR}\propto \Sigma_{\rm gas}^{1.4}$ on galactic scales. The observed cosmic SFRD, $\psi(z)$, is the total SFR summed over all galaxies per unit comoving volume, which can be recast to be the mean SFR averaged over all galaxies multiplied by the comoving number density of galaxies, $\psi(z)=\langle{\rm SFR}\rangle(z)\cdot\,n_{\rm gal}(z)$.  Integrating the Kennicutt--Schmidt relation over the total area of the galaxies leads to ${\rm SFR}\propto M_{\rm gas}^{1.4}$.  Given that the number density of galaxies has not evolved strongly within this redshift range \citep[e.g.,][]{Faber2007}, we can further relate the redshift evolution of the cosmic SFRD directly to the anticipated evolution of mean neutral gas mass over the galaxy population between different epochs, following $\langle\,{\rm SFR}\rangle\propto \langle\,M_{\rm gas}\rangle^{1.4}$. This is shown as the solid, green curve in Figure \ref{Fig:Subsamples}a.

Two key features are immediately apparent in Figure \ref{Fig:Subsamples}a.  First, the mean \ion{H}{1} content $\langle\mhi\rangle$ is found to have decreased with increasing cosmic age (or decreasing redshift) by a factor of two in low-mass galaxies of $\langle\log\,\mstar/\msun\rangle\approx 9.3$, from $\langle\mhi\rangle=(1.7\pm 0.3)\times 10^{10}\,\msun$ at $\langle\,z\rangle\approx 0.55$ to $\langle\mhi\rangle=(8\pm 2)\times 10^{9}\,\msun$ at $\langle\,z\rangle\approx 0.38$, while a more modest decline is seen in high-mass galaxies of $\langle\log\,\mstar/\msun\rangle\approx 10.6$, from $\langle\mhi\rangle=(1.3\pm 0.3)\times 10^{10}\,\msun$ to $\langle\mhi\rangle=(1.1\pm 0.2)\times 10^{10}\,\msun$ over the same period. However, due to large uncertainties, the redshift evolution of $\langle\mhi\rangle$ for both low- and high-mass galaxies agrees well with that of the cosmic star formation history. Note that a steeper redshift evolution was discussed in \citet[][their Figure~10]{Bianchetti2025NewData} for galaxies of $\log\,\mstar/\msun\approx 10$, which may be a consequence of inconsistent apertures adopted for different measurements \citep[e.g.,][]{Guo2021}. Although the uncertainties remain large, the consistent redshift evolution trend between the cosmic star formation rate density and $\langle\mhi\rangle$ in galaxies offers tantalizing support for a causal connection between neutral gas and star formation in galaxies.

One notable caveat is the dependence of the observed redshift evolution on the aperture size.  The apparent consistency in redshift evolution suggests that a substantial fraction of the measured \ion{H}{1} mass occurs outside the star-forming ISM immediate to the central galaxies (see \S\,\ref{sec:diameter}), whereas the Kennicutt–Schmidt relation is defined for the ISM rather than extended halo gas. One implication is that the accretion timescale for halo \ion{H}{1} onto the ISM may be shorter than the star-formation (depletion) timescale. To illustrate the impact of aperture choice, we present the observed redshift dependence measured with different extraction apertures in \S\ref{sec:Stacks-app}.  

In addition, the Kennicutt–Schmidt relation is defined for the \textit{total} neutral gas mass---both atomic and molecular. In our analysis, we use \ion{H}{1} as a proxy for the total neutral reservoir, implicitly assuming that the neutral phase is dominated by atomic gas traced by the 21~cm line. However, numerous studies show that the observed SFR correlates more directly with the molecular component \citep[e.g.,][]{Leroy2008}, even though star formation can proceed in predominantly atomic gas in low-metallicity environments such as outer disks and galactic halos \citep[e.g.,][]{Krumholz2012,Glover2012}. While we find consistent redshift evolution between the halo-scale mean \ion{H}{1} content and the ISM (disk-scale) SFRD of the general galaxy population, our data lack the spatial resolution to determine whether the 21~cm line-emitting gas is forming stars.

\subsection{Variation of \mhi\ with galaxy mass}
\label{sec:mevol}

An additional goal of our study is to examine how the \ion{H}{1} content varies with galaxy mass, which, in our study, is determined by \mstar. Figure \ref{Fig:Subsamples}b shows the weighted mean \mhi\ versus \mstar\ for different redshift bins.  For comparison, we also include the best-fit \mhi-\mstar\ scaling relation at $z\approx 0.36$ and its associated uncertainties from \citet{Bianchetti2025NewData}.  In addition, the dashed lines mark different constant \mhi/\mstar\ ratios.

We note two new features in Figure \ref{Fig:Subsamples}b, in addition to the trend of increasing $\langle\mhi\rangle$ with increasing redshift described in \S\ \ref{sec:zevol}.  First, with an order of magnitude difference in the mean stellar mass at both low and high redshifts, the \ion{H}{1} content remains roughly constant in these galaxies in both epochs.  The observed constant \mhi\ leads to a factor of $\approx 18$ difference in the \mhi/\mstar\ ratio between low- and high-mass galaxies, with $\mhi/\mstar\approx 6$ in galaxies of $\langle\log\,\mstar/\msun\rangle\approx 9.3$ and $\mhi/\mstar\approx 0.3$ in galaxies of $\langle\log\,\mstar/\msun\rangle\approx 10.6$ at all redshifts probed by our sample. See the last column of Table \ref{Table:Samples} for the exact values and associated uncertainties.

The observed mass dependence in our sample differs from previous studies, which report a steeper \mhi-\mstar\ correlation, specifically $\mhi\propto \mstar^{0.3}$ \citep[e.g., the results from the joint MIGHTEE-\ion{H}{1}+CHILES sample at $z\approx 0.36$;][]{Bianchetti2025NewData}.  While the $\mhi/\mstar$ ratios for the high-mass CUBS-MALS galaxies at both high and low redshifts are consistent with the MIGHTEE-\ion{H}{1}+CHILES measurements at lower redshifts, the low-mass CUBS-MALS galaxies show significantly higher $\mhi/\mstar$ ratios compared to the MIGHTEE$+$CHILES results.

\subsection{Implications and potential systematics}
\label{sec:conc}

The observed redshift evolution of the mean \ion{H}{1} gas mass in galaxies closely tracks the cosmic star formation history, supporting a scenario in which the declining availability of neutral hydrogen drives the overall decrease in the star formation rate within the general galaxy population. In contrast, previous \ion{H}{1} line-intensity mapping and damped \lya\ absorber (DLA) studies along random quasar sightlines have reported little to no evolution in the cosmological mean \ion{H}{1} mass density at $z \lesssim 1$. Although scatter remains substantial among these measurements (see e.g., \citealt{Chang2010, Neeleman2016, Walter2020} and see also \citealt{Guo2023} for a recent compilation of different measurements) and dust reddening may bias against QSO sightlines intersecting high-$N({\rm HI})$ DLAs---potentially leading to an underestimate of neutral gas mass at high redshifts \citep[e.g.,][]{Krogager2019}---the apparent lack of strong evolution in cosmological mean \ion{H}{1} mass density has been interpreted as evidence for continuous replenishment of neutral gas reservoirs through accretion or feedback processes \citep[e.g.,][]{Neeleman2016}. 

The discrepancy between the evolution of the \ion{H}{1} mass derived from our stacks and that measured by the DLA surveys suggests that these observations may probe neutral gas residing in distinct cosmological volumes.  For example, deep multi-color galaxy surveys have shown that the redshift evolution of the star formation rate per unit mass depends on galaxy mass, with more massive galaxies displaying a more rapid decline in their star formation rates over cosmic time compared to lower-mass galaxies \citep[e.g.,][]{Whitaker2014}.  Given that the CUBS-MALS galaxy sample is dominated by relatively mature, massive systems located at large angular distances from the QSO sightline (see \S\ \ref{sec:cubs} and more extensive discussion below), the steep redshift evolution observed in their mean \ion{H}{1} gas mass suggests that neutral gas reservoirs associated with mature galaxies diminish more rapidly than those around low-mass star-forming galaxies, which dominate the signals detected by line-intensity mapping and DLA surveys. 

At the same time, our findings add further complexity to the already considerable scatter seen among different stacking analyses reported in the literature. In particular, resolving discrepancies in the \(\mhi\)–\(\mstar\) scaling relation (Figure~\ref{Fig:Subsamples}b) is non-trivial, as multiple factors likely influence these results in distinct and nuanced ways.

Fundamental differences between the CUBS-MALS galaxy sample and previous MIGHTEE-\ion{H}{1}+CHILES studies include: (1) our galaxy sample is established from 11 distinct fields and therefore less susceptible to potential cosmic variance; (2) relatively mature, massive systems dominate the CUBS-MALS galaxy sample, while previous studies focus primarily on star-forming galaxies; and (3) our analysis employs a large aperture of 300 kpc in diameter for extracting the \ion{H}{1} signals, while the MIGHTEE-\ion{H}{1}+CHILES measurement is based an aperture of $\approx 100$ kpc \citep[see e.g.,][]{Bianchetti2025NewData}.

Given the prioritization of $L_*$ and $g-r>1.1$ galaxies in the CUBS redshift survey design at \(\theta \approx 3\arcmin\)–\(10\arcmin\) from the QSO sightlines (see \S\ref{sec:cubs}), a substantial fraction of the CUBS–MALS sample likely comprises more mature systems with relatively low \ion{H}{1} mass fractions. In light of differences in both sample selection and extraction aperture, the apparent agreement in the mean \ion{H}{1} mass between our high-mass subsamples and previous studies in Figure~\ref{Fig:Subsamples}b is, therefore, somewhat surprising. As shown in Figures~\ref{fig:Diameter} and \ref{Fig:Subsamples-app}, the mean \ion{H}{1} mass for the low-{\it z}, high-mass subsample measured within a 150\,kpc-diameter aperture, closer to that adopted by MIGHTEE-\ion{H}{1}+CHILES, amounts to only 55\% of the value obtained with our fiducial 300\,kpc-diameter aperture. Taken together, these considerations suggest that, once selection and aperture effects are accounted for, the intrinsic \(\mhi/\mstar\) ratios of gas-rich, star-forming galaxies are likely higher than implied by existing measurements for high-mass galaxies.

Similarly, the significant difference in the \ion{H}{1} content observed for low-mass galaxies may be attributed to aperture loss.  As demonstrated in \S\ \ref{sec:diameter} and in \S\ \ref{sec:Stacks-app}, increasing the extraction aperture diameter, from 150 kpc to 300 kpc, leads to a factor of $\approx 2$ increase in the total \ion{H}{1} flux.  Therefore, the lower \mhi\ measurements from the MIGHTEE+CHILES study (based on an extract aperture of triple the FWHM, which is $\approx 100$ kpc in diameter at $z=0.36$ for CHILES's beam as the smaller compared to that of MeerKAT) relative to ours may reflect missed 21~cm flux located beyond their extraction aperture.

In summary, we have uncovered significant \ion{H}{1} 21-cm signals extending out to a 150 kpc radius around typical galaxies at intermediate redshifts. The \ion{H}{1} mass fraction decreases with increasing stellar mass.  The redshift evolution of mean \ion{H}{1} mass, $\langle\mhi\rangle$, in both low- and high-mass field galaxies, aligns well with the expectation from the cosmic star formation history.  While the inferred \mstar\ correlates strongly with the observed\ $g-r$ color, the large scatter seen in Figure \ref{fig:RedshiftStellarMass} also suggests the presence of a wide range in the star formation histories of individual galaxies \citep[see e.g.,][]{Bell2003}. A natural next step would be to investigate how the observed \ion{H}{1} content depends on the galaxy's star formation history to gain a deeper understanding of the discrepancies reported in the literature. The \ion{H}{1} mass measurements based on 21-cm absorption and emission signals from absorption-selected galaxies may also provide clues to the discrepancy concerning constraints from the DLAs.

\begin{acknowledgements}
We thank an anonymous referee for their constructive comments that helped improve the presentation of this work.  We acknowledge the Massachusetts Institute of Technology's Office of Research Computing and Data for managing the Engaging cluster computer on which this letter's research was performed.
This work is based on observations made with ESO Telescopes at the Paranal Observatory under program ID 0104.A0147(A), observations made with the 6.5-m Magellan Telescopes located at Las Campanas Observatory, spectroscopic data gathered under the HST-GO-15163.01A program using the NASA/ESA Hubble Space Telescope operated by the Space Telescope Science
Institute and the Association of Universities for Research in Astronomy, Inc., under NASA contract NAS 5--26555, and imaging spectroscopic data from the MeerKAT telescope.  MeerKAT is operated by the South African Radio Astronomy Observatory, which is a facility of the National Research Foundation, an agency of the Department of Science and Innovation. 
The MeerKAT data were processed using the MALS computing facility at IUCAA (https://mals.iucaa.in/releases).
The Common Astronomy Software Applications (CASA) package is developed by an international consortium of scientists based at the National Radio Astronomical Observatory (NRAO), the European Southern Observatory (ESO), the National Astronomical Observatory of Japan (NAOJ), the Academia Sinica Institute of Astronomy and Astrophysics (ASIAA), the CSIRO division for Astronomy and Space Science (CASS), and the Netherlands Institute for Radio Astronomy (ASTRON) under the guidance of NRAO.
The National Radio Astronomy Observatory is a facility of the National Science Foundation operated under cooperative agreement by Associated Universities, Inc.
NG acknowledges the NRAO for its generous financial support for the sabbatical visit to Socorro, during which part of this work was conducted.
HWC acknowledges partial support from HST-GO-17517.01A and NASA ADAP 80NSSC23K0479 grants.
EB acknowledges support from NASA under award number 80GSFC24M0006.
SC gratefully acknowledges support from the European Research Council (ERC) under the European Union’s Horizon 2020 Research and Innovation programme grant agreement No 864361.
MCC is supported by the Brinson Foundation through the Brinson Prize Fellowship Program.
CAFG was supported by NSF through grants AST-2108230 and AST-2307327; by NASA through grants 21-ATP21-0036 and 23-ATP23-0008; and by STScI through grant JWST-AR-03252.001-A.
SL acknowledges support by FONDECYT grant 1231187.
\end{acknowledgements}

\facilities{Blanco (DECam), Magellan:Baade (FourStar, IMACS), Magellan:Clay (LDSS-3), MeerKAT, VLT:Yepun (MUSE)}

\software{
ARTIP \citep{Gupta21},
Astropy \citep{Astropy22},
CASA \citep{TheCASATeam2022CASAAstronomy},
katbeam \citep{deVilliers2022MeerKATBand},
Matplotlib \citep{Hunter07},
NumPy \citep{Harris20},
Regions \citep{Bradley2022Astropy/regions},
SciPy \citep{Virtanen2020SciPyPython},
and Spectral-cube \citep{Ginsburg2015RadioMore}.
}

\bibliographystyle{aasjournal}
\bibliography{stack}

\appendix
\counterwithin{figure}{section}

\section{Effects and implications of changing aperture sizes on the extracted \ion{H}{1} signals}
\label{sec:Stacks-app}

As discussed in Section~\ref{sec:diameter}, the nominal aperture size for extracting the \ion{H}{1} signal is not known a priori. In principle, robust detection of \ion{H}{1} emission depends on matching the aperture size to the spatial extent of the gas. A fundamental limitation in our analysis is set by the telescope’s synthesized beam, which corresponds to a diameter of 150 kpc at the maximum redshift $z = 0.634$ in the CUBS–MALS sample. We therefore explore a range of extraction apertures in multiples of 150 kpc. Here we present spectral stacks and the inferred mean \ion{H}{1} mass of CUBS galaxies using apertures with diameters of \qty{150}{kpc} and \qty{450}{kpc}.

\begin{figure}[h]
\centering
    \includegraphics[width=0.45\textwidth]{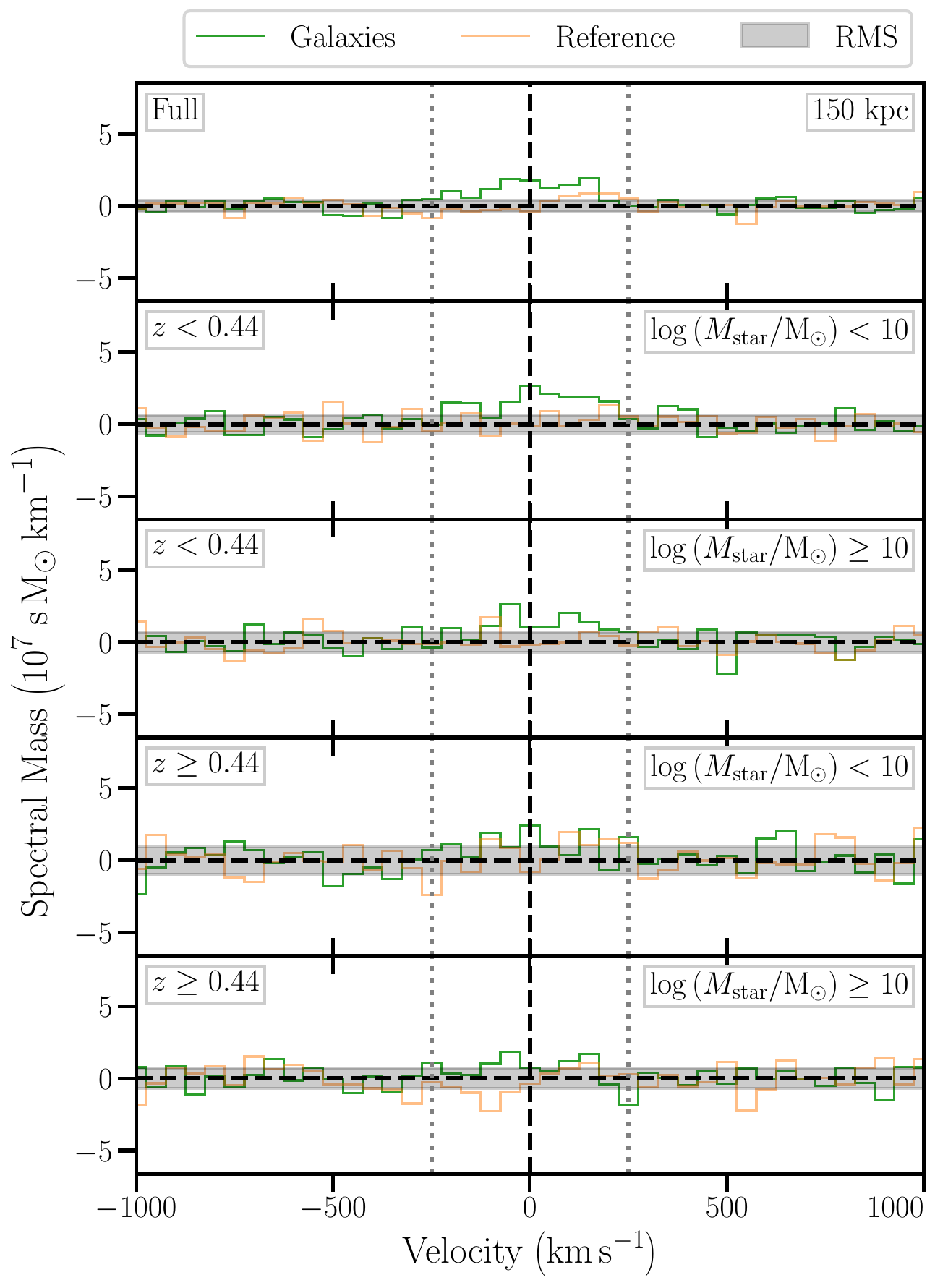}
    \includegraphics[width=0.45\textwidth]{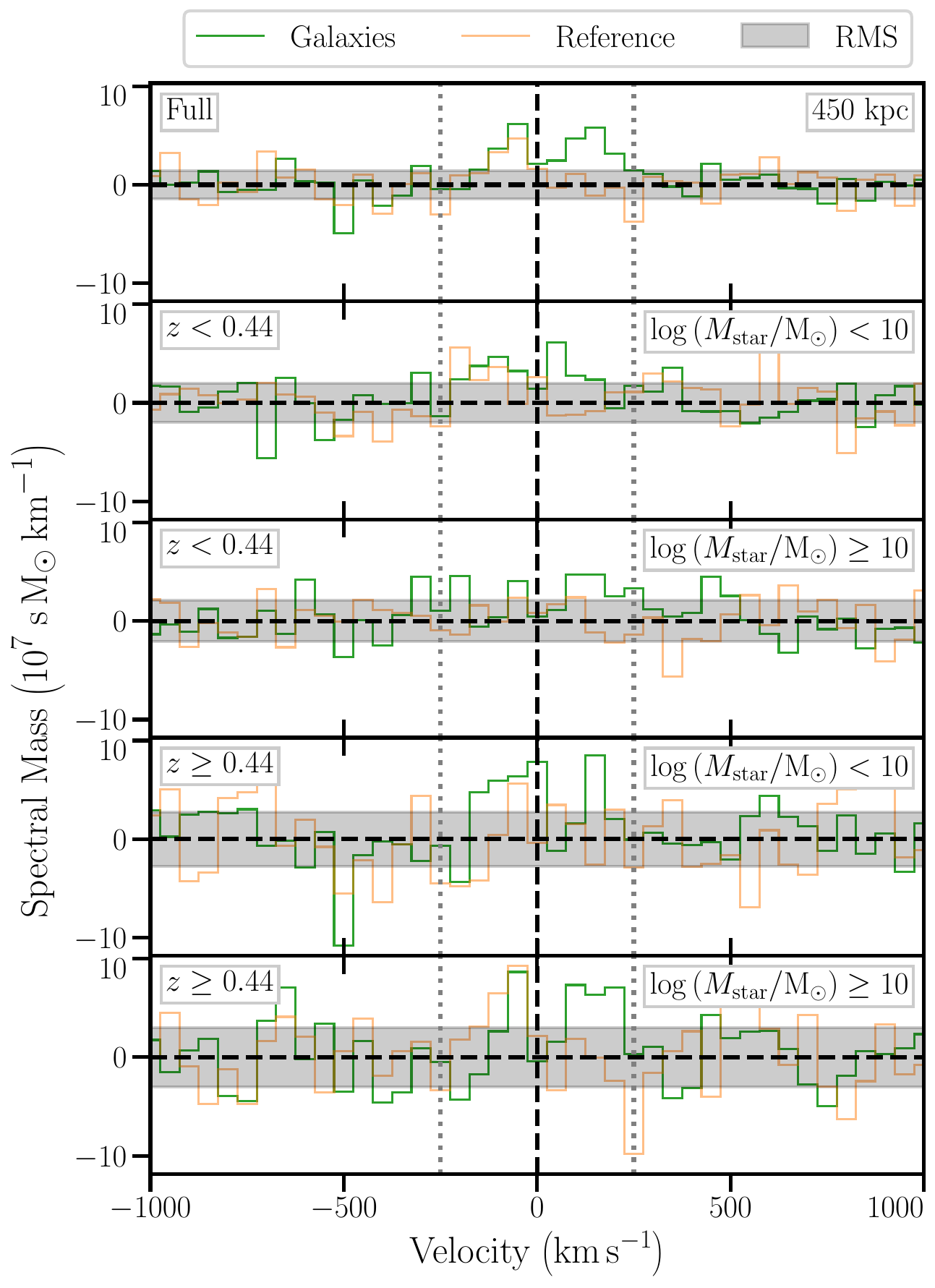}
    \caption{Weighted mean spectral stacks for apertures with diameters of \qty{150}{kpc} ({\it left}) and \qty{450}{kpc} ({\it right}).  The remaining details are the same as in Figure \ref{Fig:Stacks}.
    }
    \label{Fig:Stacks-app}
\end{figure}

\begin{figure*}
\centering
    \includegraphics[width=0.45\textwidth]{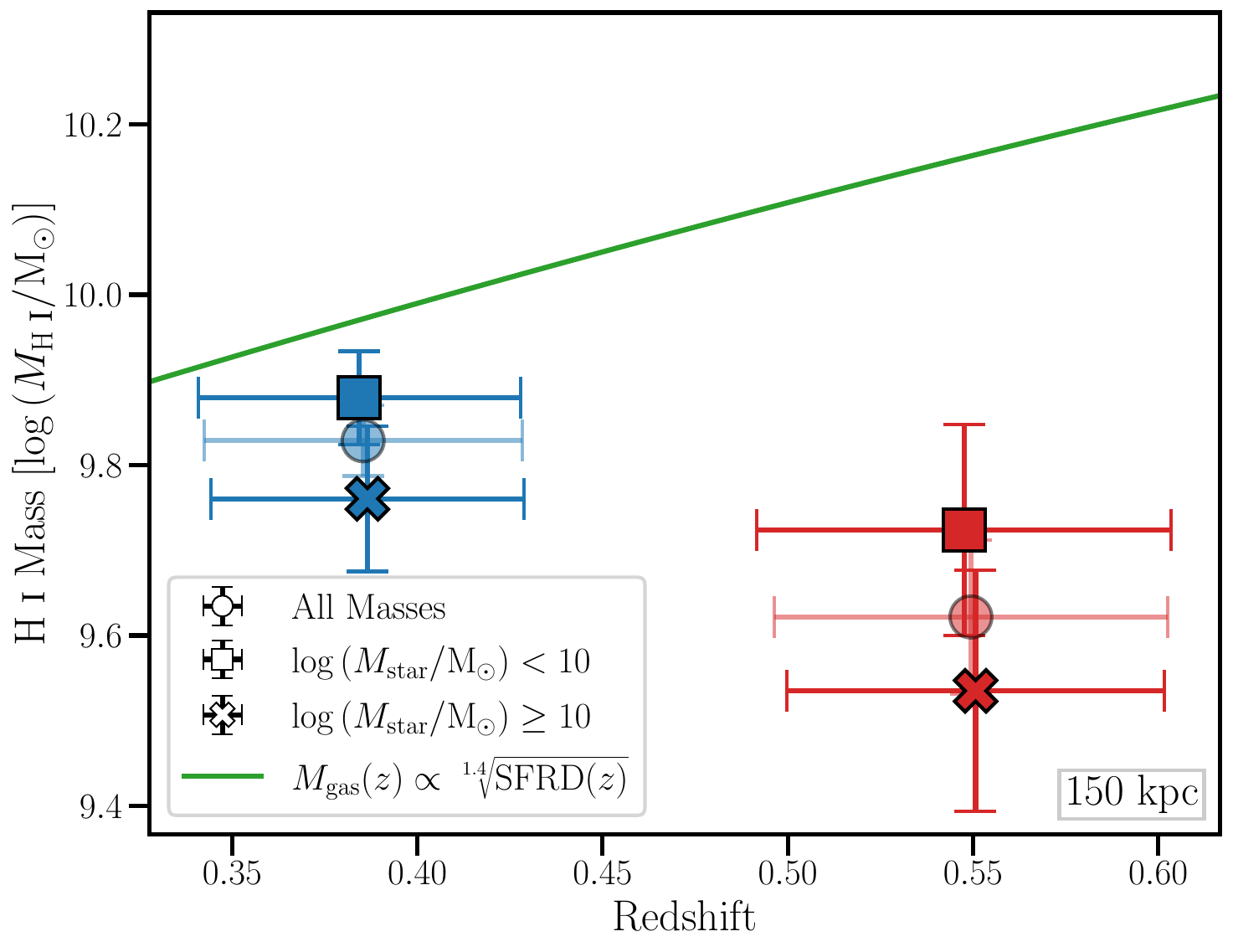}    \includegraphics[width=0.45\textwidth]{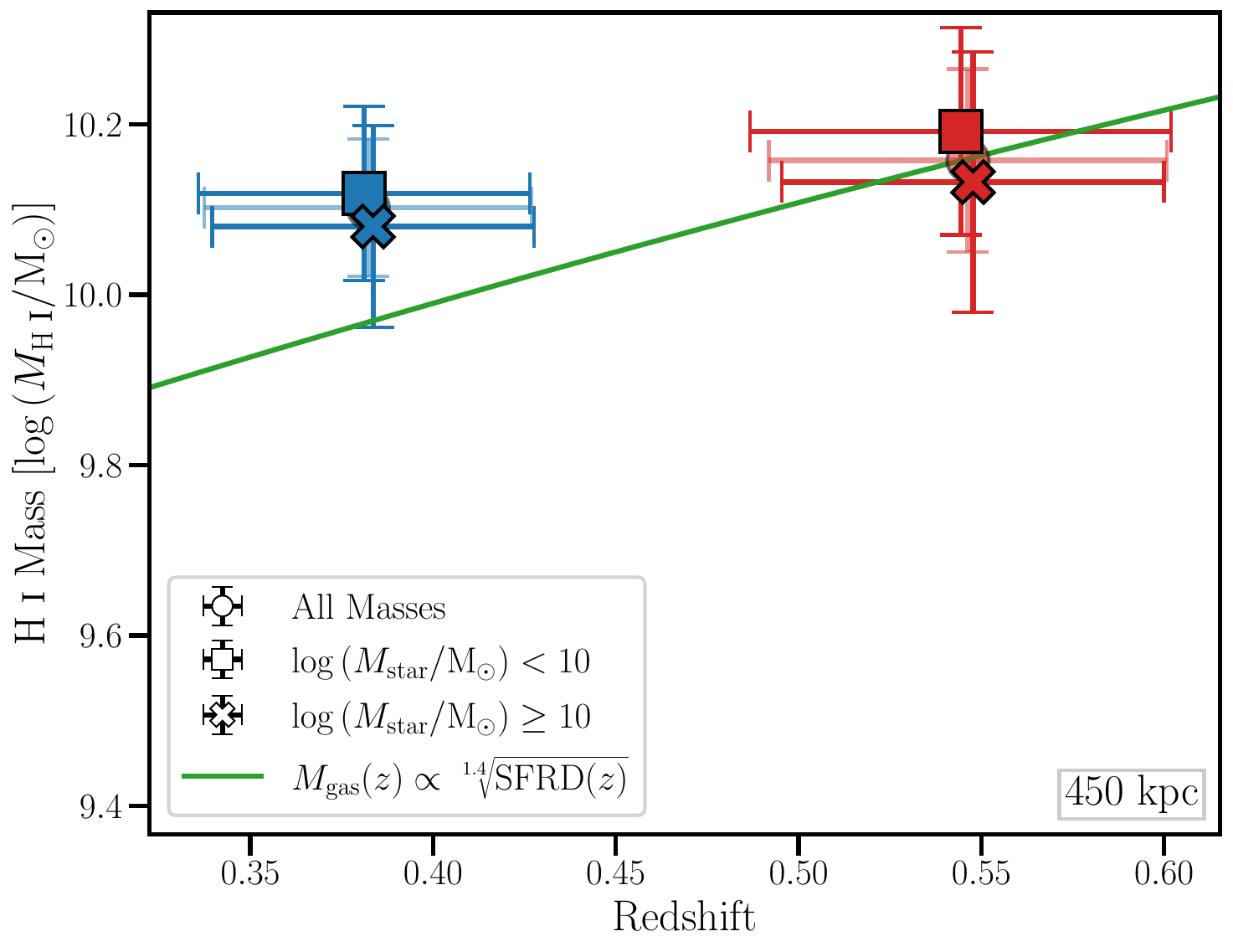}
    \caption{Redshift evolution of measured mean \ion{H}{1} mass for apertures with diameters of \qty{150}{kpc} ({\it left}) and \qty{450}{kpc} ({\it right}). The remaining details are the same as in Figure \ref{Fig:Subsamples}.}
    \label{Fig:Subsamples-app}
\end{figure*}

It is clear from comparing the stacked spectra between Figures \ref{Fig:Stacks} and \ref{Fig:Stacks-app} that doubling the extraction aperture size, from 150 kpc to 300 kpc, increases the extracted total \ion{H}{1} 21~cm line flux by a factor of more than two.   Further increasing the aperture diameter to 450 kpc results in a substantial rise in noise without a corresponding gain in signal.  At the same time, Figure \ref{Fig:Subsamples-app} shows that the adopted aperture size also has a dramatic impact on the inferred redshift evolution of mean \ion{H}{1} mass.  In contrast to the increasing mean \ion{H}{1} mass with increasing redshift using a 300-kpc diameter apture, the inferred mean \ion{H}{1} mass within a 150-kpc diameter aperture, decreases with increasing redshift.  However, as described in \S\,\ref{sec:mals}, the synthesized beam has ${\rm FWHM}\approx 11\arcsec$ corresponding to 50 kpc at $z\approx 0.3$ and ${\rm FWHM}\approx 22\arcsec$ corresponding to 150 kpc at $z\approx 0.6$.  A 150-kpc diameter aperture therefore corresponds to $3\times{\rm FWHM}$ for low-$z$ galaxies, but merely $1\times {\rm FWHM}$ for the high-$z$ subsample.  The apparent trend in Figure \ref{Fig:Subsamples-app}, therefore, needs to be corrected for differential aperture loss.

We therefore conclude that an extraction aperture of 300 kpc in diameter provides the optimal balance for maximizing the \ion{H}{1} signal while minimizing noise and systematic biases.  In addition, the substantial increase in the total 21~cm line flux from 150 kpc to 300 kpc-diameter aperture has also revealed the presence of substantial \ion{H}{1} gas at distances far beyond the stellar body around these galaxies.

\end{document}